\documentclass[%
reprint,superscriptaddress,groupedaddress,unsortedaddress,runinaddress,frontmatterverbose, 
 amsmath,amssymb,
aps,
prx,
floatfix,
longbibliography,
]{revtex4-2}

\usepackage[pdftex]{graphicx}
\usepackage{float}
\usepackage{dcolumn}
\usepackage{bm}

\usepackage[english]{babel}
\usepackage[utf8]{inputenc}
\usepackage[colorinlistoftodos, color=green!40, prependcaption]{todonotes}
\usepackage[pdftex,colorlinks=true,allcolors=black,pdftitle={Article},pdfauthor={Author}]{hyperref}
\usepackage{natbib}

\begin{document}
\title{Momentum-resolved resonant photoelectron spectroscopic study for 1T-TiSe$_2$: Observation of  negative $q$ in the Fano resonance due to inter-atomic interaction in the valence band}

\author{Shin-ichiro Tanaka}
    \email[Correspondence email address: ]{stanaka@sanken.osaka-u.ac.jp}
    \affiliation{SANKEN, The Institute of Scientific and Industrial Research, Osaka University, Ibaraki 567-0047, Japan }
\author{Shigemasa Suga}%
\affiliation{SANKEN, The Institute of Scientific and Industrial Research, Osaka University, Ibaraki 567-0047, Japan }
\author{Keiji Ueno}%
\affiliation{Graduate School of Science and Engineering, Saitama University, Saitama 338-8570, Japan}
\author{Keisuke Fukutani}%
\affiliation{Institute for Molecular Science, Myodaiji, Okazaki 444-8585, Japan}
\author{Fumihiko Matsui}%
\affiliation{{UVSOR Synchrotron Facility, Institute for Molecular Science, Myodaiji, Okazaki 444-8585, Japan}}

\date{\today} 

\begin{abstract}
The remarkable properties of (1T-)TiSe$_2$ among the transition metal dichalcogenides have attracted the attention of many researchers due to its peculiar behavior during the charge density wave (CDW) transition. Therefore, it is highly desirable to study its electronic structure down to the atomic orbitals.
In the present research, we applied momentum-resolved resonant photoelectron spectroscopy to study TiSe$_2$ at the Ti2p$\rightarrow$Ti3d absorption edge by using  a momentum microscope, which can simultaneously detect the electronic states in a wide ($k_x,k_y$) range.
We have also used constant initial state (CIS) spectroscopy and density functional theory (DFT) calculations to reveal the hybridization between the Ti3d and Se4p orbitals within the valence band at the $\Gamma$ point at room temperature.
In addition, an interesting result comes from our analysis of the CIS spectrum for the energy band located at a binding energy of 2 eV at the M-point. 
This band, mainly composed of the Se4p orbital, exhibited a Fano line profile at the Ti2p edge, with a negative value of the parameter `$q$’.
This is the first clear evidence of the inter-atomic interaction during the valence band photoelectron emission process.
This behavior differs significantly from the standard resonant photoelectron emission, which usually involves intra-atomic interactions.
It also differs from the multi-atom resonant photoelectron emission (MARPE) observed in the core-level photoelectron emission, as we focus on the photoelectron emission from the valence band in this research.
\end{abstract}


\maketitle

\section{Introduction}

\begin{figure*}
  \centering
\includegraphics[width=16cm]{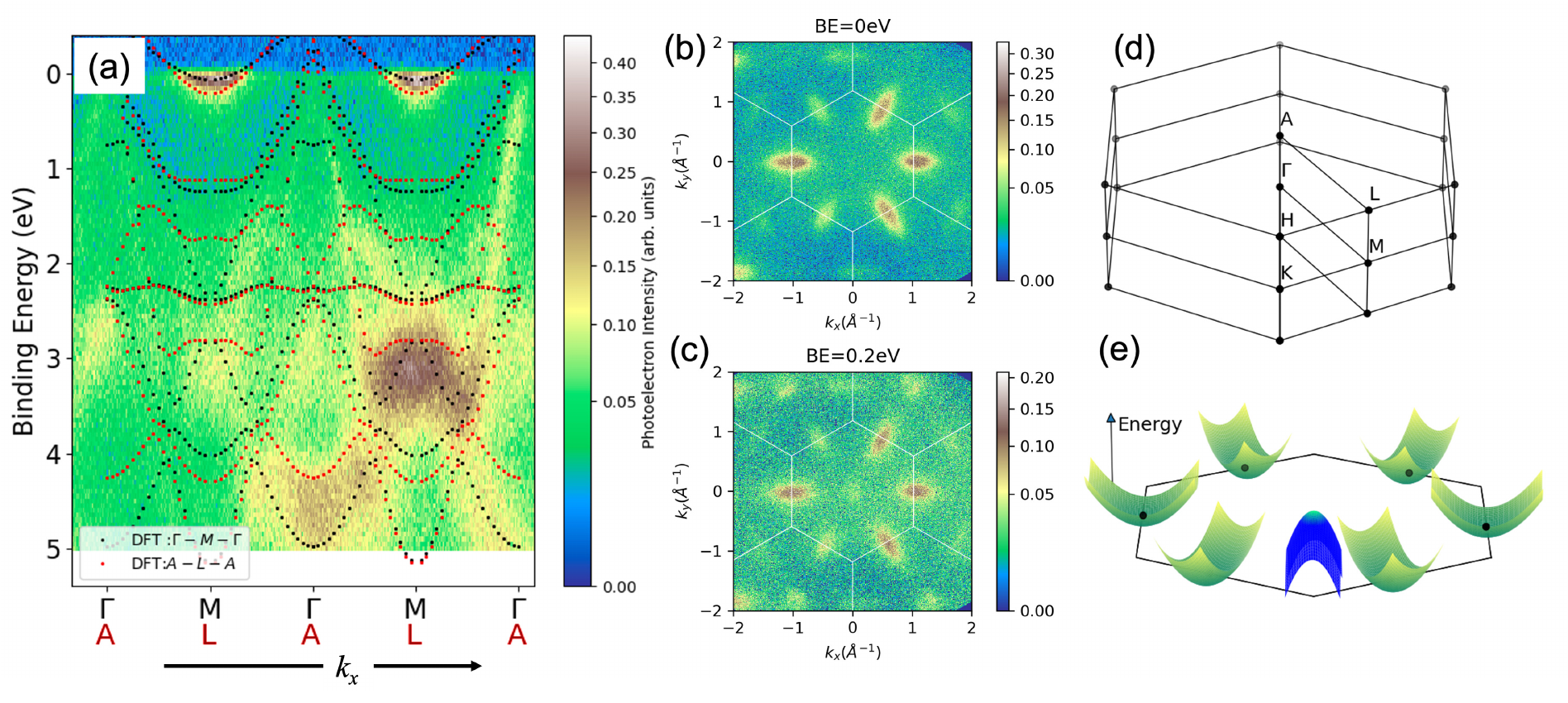}
  \caption{ \label {Fig1}
  (a) Calculated band dispersions of TiSe$_2$ along the $\Gamma$-M line (black dots) and the A-L line (red dots) together with the experimentally obtained photoelectron intensity map in a color scale shown at the right side of the panel taken at photon energy ($h\nu$) of 80 eV and at room temperature.
   (b,c) Iso-energy photoelectron intensity maps at the binding energy (BE) of 0 and 0.2 eV, respectively,  under the same condition as (a).  
   (d) the 3D Brillouin zones and high symmetry points. (e): Schematic of the electronic structure of TiSe$_2$ near the Fermi level, where 
   the electron  and hole pockets are located at M (L)  and $\Gamma$ (A) points, respectively.}
\end{figure*} 

The layered  transition-metal-dichalcogenide (TMDC) TiSe$_2$ has been extensively studied for decades 
due to its characteristic charge-density-wave (CDW) transition behavior~\cite{Gruner1988}.
The $(2\times 2\times 2)$ superstructure, which forms below a critical temperature ($T_{\rm c}$) of approximately 200 K, was first reported by Di Salvo {\it et al.}~\cite{DiSalvo1976}. 
Debates have arisen concerning the driving force behind the CDW transition. 
One perspective suggests the involvement of the electron-phonon interaction~\cite{Clandra2011,Wegner2020}, 
 while another proposes the exciton condensation as the driving factor~\cite{Cercellier2007}. 
 Although the latter perspective is currently considered more favorable, especially in light of critical 
 measurements using electron energy loss spectroscopy~\cite{Kogar2017}, 	
the origin and nature of the CDW transition in TiSe$_2$ remains controversial. 

The band structures, derived from our density functional theory (DFT) calculations, are presented along the M-$\Gamma$-M line (represented by the black dotted line) and the A-L-A  line (depicted by the red dotted line) in Fig.~1(a) along with the angle/momentum-resolved photoelectron spectroscopy (ARPES/MRPES) intensity map (recorded at room temperature with $h\nu$=80 eV) . The notation of the symmetry points in the Brillouin zone is illustrated in Fig.~1(d). Further details about the measurements and calculations will be provided later.

Note that the $k_z$ value  in the ARPES/MRPES map, calculated  assuming the inner potential of 13 eV~\cite{Chen2016}, 
varies from $-0.08$ ${\rm \AA}^{-1}$ (BE=5 eV) to 0.06 ${\rm \AA}^{-1}$ ($E_{\rm F}$) at  $\Gamma (A)$  , and $-0.18$ ${\rm \AA}^{-1}$ (BE=5 eV)  to $-0.06$ ${\rm \AA}^{-1}$ ($E_{\rm F}$) at M(L). 
The bands accountable for the CDW transition are schematically illustrated in Fig.~1(e). They consist of the electron pocket (often referred to as the conduction band) situated at the M(L) point, clearly distinguishable in Fig.~1(b), and the hole pocket (often referred to as the valence band) located at the $\Gamma$(A) point. The valence band at the $\Gamma$ (A) point is also observable in the ARPES/MRPES map  but slightly weaker than the electron pocket in Fig.~1(c).
In the proposed exciton condensation process~\cite{Cercellier2007},  thermal excitations lead to the creation of electrons in the conduction band and holes in the valence band, providing the electron-hole bound states known as excitons. 
With a periodicity determined by the spanning vector that connects the valence band maximum to the conduction band minimum, this  can cause a transition to a coherent ground state of condensed excitons and the superlattice formation~\cite{Cercellier2007}.  
Consequently, understanding the orbitals of these bands is critical to a deeper understanding of the CDW transition.
In the simplest description, the electronic configuration of TiSe$_2$ is (Se4p)$^6$(Ti3d)$^0$, 
where the valence band is commonly assumed to be derived from the Se4p orbital and the conduction band from the Ti3d orbital. 
However, some studies~\cite{Stoffe1985,Rossnagel2002,Hellgren2017,Watson2019} have proposed a hybridization between the Ti3d and Se4p orbitals in the valence band located near the $\Gamma$ point in the CDW phase. Therefore, an experimental investigation is desirable to elucidate the orbital character of these electronic states.

	In order to access to this objective, resonant photoelectron spectroscopy utilizing synchrotron radiation is an ideal tool and has been extensively employed to probe the electronic structure of various materials~\cite{Davis1986,Kotani2001,FADLEY20102}. In resonant photoelectron emission,  the photon energy is scanned around the core-to-bound excitation energy, leading to multiple pathways of excitation and decay. 
 In instances where the incident photon generates a core hole (in the current case, it is Ti2p) and the bound electron in the conduction band (illustrated as  A in Fig. 2), subsequent exchange and direct Auger decay~\cite{RevModPhys.44.716}  (denoted as B and B' in Fig. 2) result in a hole within the valence band.

\begin{figure}[bthp]
\includegraphics[width=8cm]{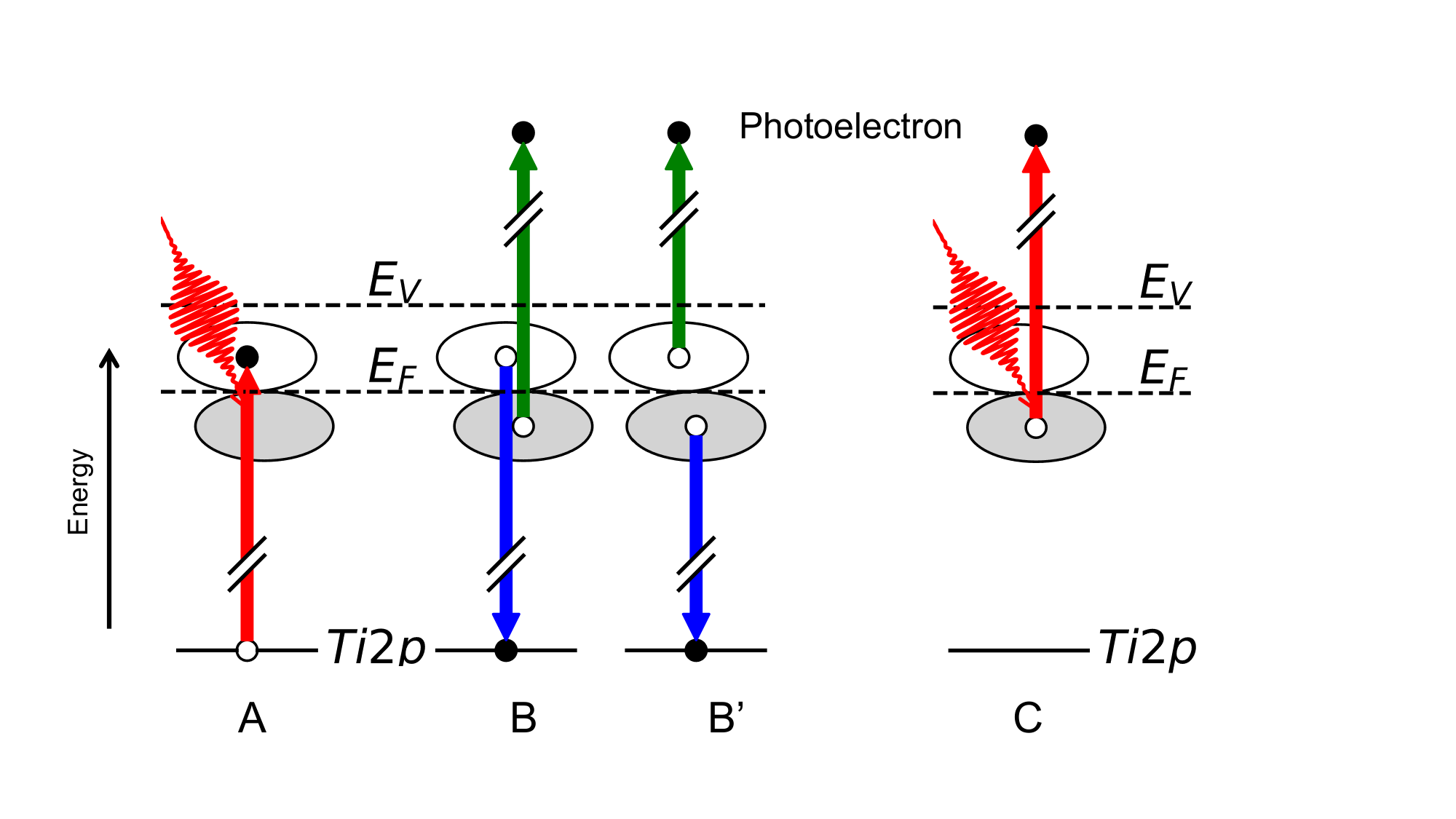}
  \caption{ \label {Fig2} The schematic model during the photoelectron emission process in the resonant photoelectron spectroscopy. See the text for details.}
\end{figure}

 The notations of the ``exchange" and ``direct" are according to Bambynek {\it et al.}~\cite{RevModPhys.44.716}
Since  the final state of this process is identical to that of the normal (direct) photoelectron emission  from the valence band, shown as C in Fig.~\ref{Fig2}, the photoelectron intensity from the valence band is  resonantly enhanced in the vicinity of the core-level photo-absorption energy. 
Resonant enhancement in the photoelectron emission is more pronounced when the interaction between the core-hole decay and the Auger-electron 
emission is stronger. 
This interaction exhibits significantly greater strength when the relevant core and valence/conduction bands are contributed from the atomic orbitals of the same atom, which is known as intra-atomic Auger decay, as opposed to when they are contributed by different atoms (inter-atomic Auger decay). As a result, conducting resonant photoelectron spectroscopy at the Ti2p excitation allows us to distinguish the Ti3d-derived band from the Se4p-derived band.
Although some studies using resonant photoelectron spectroscopy on the Ti2p excitation in TiSe$_2$ have been reported ~\cite{Shkvarin2012,Chuang2020}, the measurements were momentum-integrated and the bands near the Fermi level were not well resolved. 
Therefore, a detailed study  of the momentum-resolved resonant photoelectron spectroscopy on TiSe$_2$ is highly desirable.
Here we adopt a new approach, where the momentum-resolved CIS spectra are obtained and compared with the momentum-resolved DOS of specific atomic orbitals.

Fano resonance has been widely investigated in may spectroscopic fields so far~\cite{Fano1961,Limonov2021} including the photoelectron spectroscopy~\cite{Davis1986,Kakizaki1983,Kotani2001}. 
In the resonant photoelectron spectroscopy, it arises from interference between distinct energy levels during excitation, followed by the Auger decay (process A→B/B'), and excitation into continuous levels (process C). Constant initial-state spectroscopy (CIS), in which the photoelectron intensity at a specific binding energy of interest is plotted as a function of photon energy, has been used to examin the Fano shape. 
The Fano resonance~\cite{Fano1961,Limonov2021} is formulated as 
\begin{equation}
\label{Fano_formula}
I=\frac{I_0}{1+q^2}\frac{(q+\epsilon)^2}{1+\epsilon^2}
\end{equation}
, where $q$ is the asymmetry parameter that determins the Fano-type spectral shape, $I_0$ is the intensity, and $\epsilon$ is the normalized energy as 
\begin{equation}
\label{Fano_energy}
\epsilon=	\frac{x-x_0}{\Gamma}
\end{equation}
, where $x$ is the photon energy, $x_0$ is the core-absorption energy,  $\Gamma$ is the Lorentzian width.
With increasing $|q|$, the peak shape becomes more symmetric and 
ultimately becomes a Lorentzian curve when $|q|=\infty$, implying  that the whole process is dominated by the distinct excitation 
($A \rightarrow B/B’$). 
In  real cases, both the continuous and distinct excitation are involved, and finit  postive $q$'s  have been observed
reflecting the relative strength and phase of the two excitations.~\cite{Davis1986,Kakizaki1983,Kotani2001} 

Meanwhile, it has been highlighted that resonant photoelectron emission can arise from the inter-atomic Auger process. 
When a core-level electron of one atom is excited into an empty state at a higher energy than the Fermi level, 
it may be possible to influence the neighboring atoms when they are bonded directly. 
Then, at the photon energy corresponding to the absorption edge of one atom, 
the photoelectron emission from the  the neighboring atoms may be modulated via the inter-atomic interaction between two atoms due to their direct bonding. 
This phenomenon is commonly referred to as multi-atom resonant photoemission (MARPE)~\cite{Kay1998, PhysRevB.63.115119, FUJIKAWA2022147202}.
The first reported case was the resonant behavior of the O1s core-level photoelectron emission at the
Mn2p absorption edge in MnO~\cite{Kay1998}.
Subsequently, MARPE phenomena have been claimed in various materials in the context of the core-level photoelectron emission~\cite{KIKAS2000275, Mannella2006, Kuznetsova2019}, although an ongoing debate persists regarding the observability of MARPE ~\cite{Finazzi2000, Nordlund2001, GAO200211}.
In the context of valence band photoelectron emission, however,
the resonant effect  could easily be overshadowed by conventional resonant photoelectron emission 
because the inter-atomic interaction is significantly weaker than the intra-atomic interaction.
Although  MARPE  associated with the photoelectron emission from the core level has been reported,  
we believe that no MARPE from the valence band has been proved so far. 
It should be noted that inter-atomic Auger electron decay involving  the valenve band has been suggested in TiO$_2$ based on the
experimental results in electron-stimulated  ion desorption,
where the electron threshold energy for O$^+$ ion desorption coincided with the Ti3p excitation~\cite{KnotekFeibelman1978}.
However, according to a later work on TiO$_2$ using electron-ion coincidence spectroscopy, which is a more sophisticated and 
specific experimental technique, along with systematic comparisons of ion desorption from other metal oxides, 
 their proposal was called  into question~\cite{TANAKA200443}.
Therefore, we examined the possibility of the electron emission from the valence band due to the inter-atomic 
interaction for the first time by using the momentum-resolved experiment by photoelectron momentum microscopy (PMM).

\section{Details of Experiments and Calculations}

All the experiments were carried out at the soft X-ray beamline BL6U of the UVSOR synchrotron radiation facility of
Institute for Molecular Science (IMS), Okazaki, Japan. The endstation of the beamline consists of a
load-lock chamber, a sample-prepare chamber and a measurement chamber which 
haused a momentum microscope with a single hemispherical electron energy analyzer  manufactured by SPECS Surface Nano Analysis GmbH, which enabled
us to achieve an Angle/Momentum-Resolved PhotoElectron Spectroscopy(ARPES/MRPES)~\cite{SugaSekiyama,Matsui_2020,MatsuiRSI2023}. 
The grazing-incidence monochromator  with a varied-line-spacing plane grating (photon energy range was 40-700 eV) was used.
The photon is always horizontally polarized,
corresponding to the p-polarization with respect to the sample surface. 
The momentum microscope was set as its objective lens axis 	 is normal to the sample surface.

1T-TiSe$_2$ single crystals were grown by the chemical vapor transport method~\cite{Lieth1977,Ueno2015}. 
First, stoichiometric amounts of Ti powder and Se lumps, and iodine as a transport agent were vacuum-sealed in a quartz ampoule. 
Next, the quartz ampoule was placed in a horizontal tube furnace and heated to 900$^{\circ}$C in the raw material zone and 800$^{\circ}$C in 
the crystal growth zone for 1 week for the CVT growth. 
Then, the sample was taken out from the quartz ampoule, and inserted into the ultra-high vacuum condition.
The sample was cleaved to obtain an atomically flat clean surface just before the measurements, which typically lasted less than 12 hours. 
The condition of the sample surface was
checked by the ARPES/MRPES and core-level spectroscopy which showed no trace of any contamination.

The DFT calculation was carried out by using the open-source Quantum Espresso package (ver. 6)~\cite{Giannozzi_2009,Giannozzi_2017}
installed at  Research Center for Computational Science in IMS, Okazaki, Japan.
The pseudopotential files for Ti and Se were  \verb" Ti.rel-pbe-spn-kjpaw_psl.1.0.0.UPF" and \verb" Se.rel-pbe-n-kjpaw_psl.0.2.UPF",
 both of which
 were projector-augmented wave (PAW) type and generated using the “atomic” code produced by A. Dal Corso  through  fully-relativistic calculation. The energy cutoff of the plane wave and charge density were set to 52.0 Ry (707 eV) and 576 Ry ( 9197 eV), respectively. Before calculating the self-consistent electron density, the crystal structure was optimized using the variable cell optimization (vc-relax) method, 
 which resulted in the  optimized lattice parameters as $a=b=3.544$, and $c=6.693$ ${\rm \AA}$. 
 The self-consistent field calculation was followed by a non-self-consistent field calculation using dense $64 \times 64 \times12 $ $k$-meshes 
 in the Brillouin zone. Next, the results were used to determine the band energies, density of states, and projected density of states, wherein the contribution of each individual atomic orbital was indicated.

\section{Results}

\begin{figure}[tbhp]
  \centering
 
\includegraphics[width=8cm]{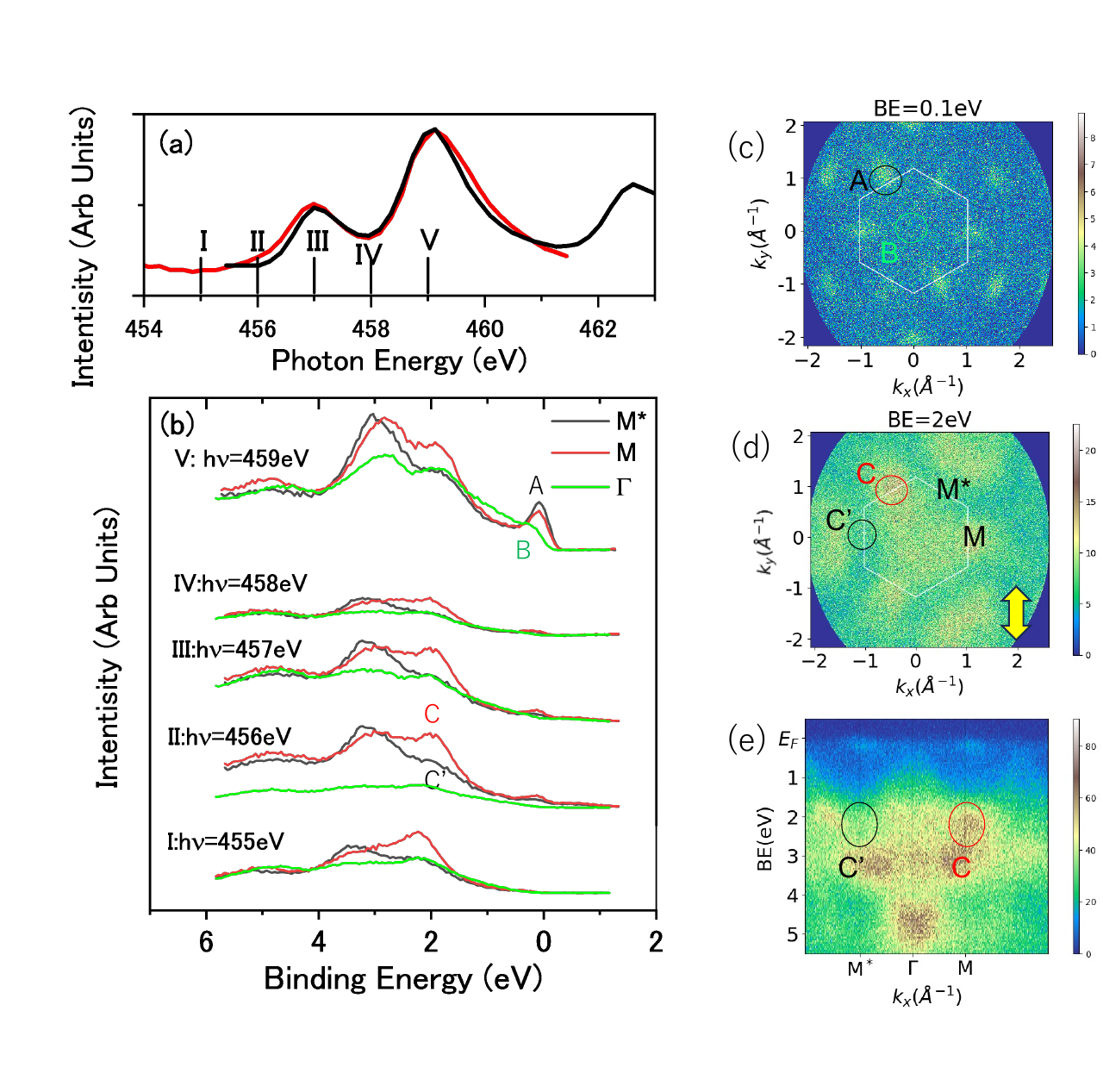}
  \caption{  \label {Spectr1} (a) The Ti-$LVV$ Auger electron (black line) and valence band photoelectron (red line) yields of TiSe$_2$ as functions of the photon energy at room temperature.
  (b) Photon-energy dependent  photoelectron spectra at M$^*$, M and $\Gamma$ points. The  used photon energies $I$-$V$  are given in 
  Fig.~\ref{Spectr1}(a).  (c,d)  Iso-energy photoelectron intensity maps at BE=0.1 eV and 2 eV, respectively, 
  (e) the photoelectron  intensity map along the M$^*$-$\Gamma$-M line.
  }
\end{figure}

\begin{figure}[tbhp]
  \centering
\includegraphics[width=8cm]{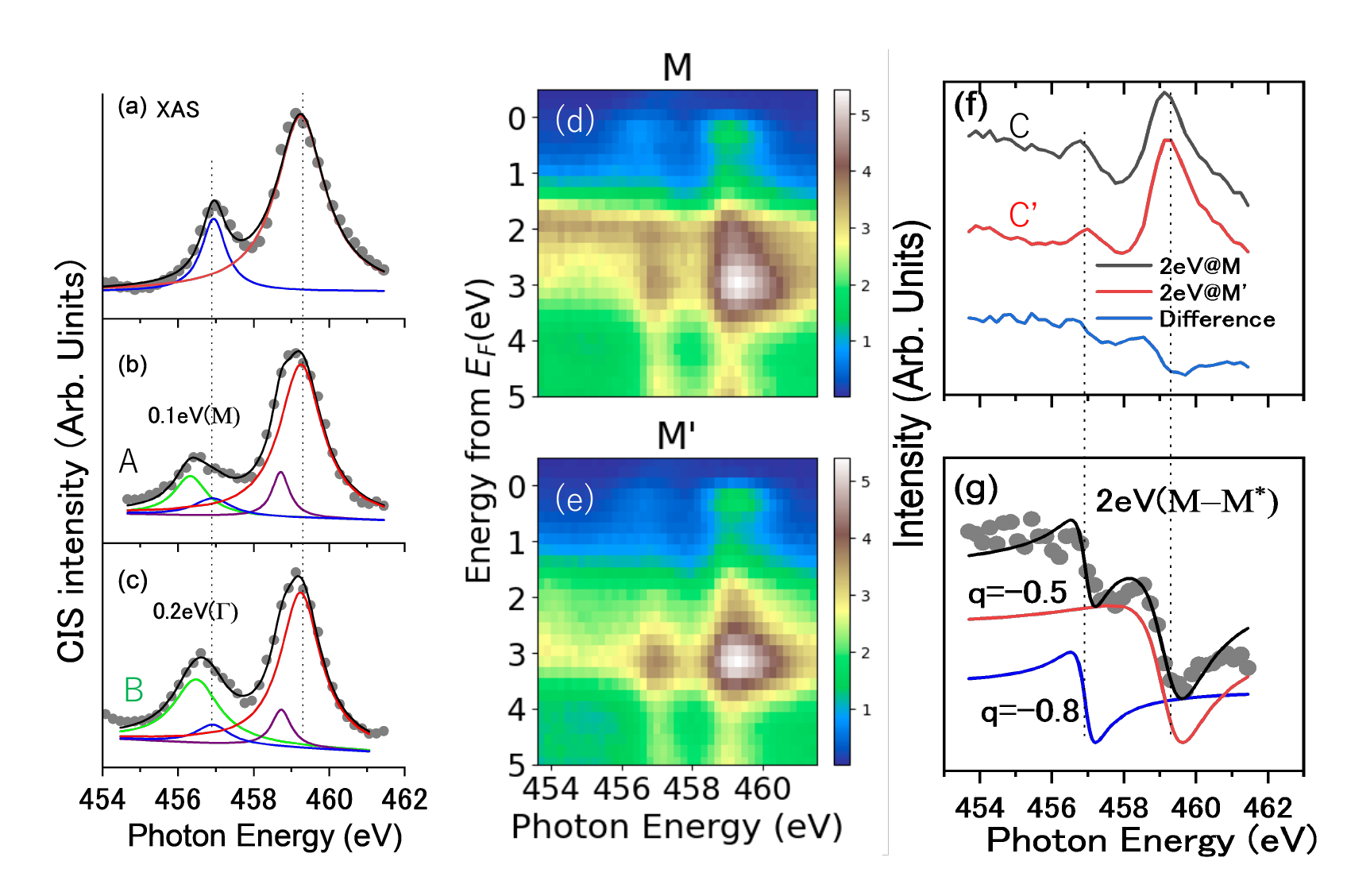}
  \caption{  \label {Spectr2} 
  (a) Ti2p XAS spectra obtained by means of the  photoelectron-yield of the entire valence band (same as the spectrum in black line in Fig.~\ref{Spectr1}(a)) (dots), 
  and the result of fitting using two Lorentzian curves and linear background. 
  (b,c) Constant Initial State (CIS) spectra (dots) and
  the result of fitting using four Lorentzian curves and linear background for the BE=0.1 eV at the M point (peak A defined in Fig.~\ref{Spectr1}(b) ) and 
  BE=0.2 eV at the $\Gamma$ point (peak B defined in Fig.~\ref{Spectr1}(b) ), respectively.
  The peak height of the XAS and CIS spectra are normalized to the same value from (a) to (c). 
  (d,e) Photoelectron intensity maps as functions of the binding energy and the photon energy at the M and M$^{*}$ points, respectively.
  (f) CIS spectra for BE=2 eV at M point (peak C defined in Fig.~\ref{Spectr1}(b) ; red line) and M$^{*}$ point (peak C' defined in Fig.~\ref{Spectr1}(b); black line) and their difference (blue line). The intensities for the peaks C and C' are averaged among three equivalent  M and M$^*$ points, respectively.
  (g) the difference in the CIS intensity between the peaks  C and C' (dots) together with the 
  result of the least-square fitting using two Fano resonance curves with $q=-0.5$ and $-0.8$ (lines) .
  }
\end{figure}

Figure \ref{Spectr1}(a) shows the electron yield curves at the Ti-$L_3$ edge, 
acquired using Ti $LVV$-Auger electron yield (solid black line) and photoelectron yield (solid red line) from the entire valence band (VB), 
where the binding energy and and $|k_{||}|$ are set within 5 eV and 5${\rm \AA}^{-1}$, respectively, are displayed. 
These intensities were normalized to align the background and peak intensities. 
Notably, these yields closely resemble the X-ray absorption spectra of TiSe$_2$ previously reported~\cite{Pal2019}. The discernible peaks at 457 (III) and 459 eV (V) are attributed to the excitations from Ti2p$_{3/2}$ to Ti3d-$t_{2g}$ and Ti3d-$e_{g}$ levels, respectively~\cite{Pal2019}.
Figure \ref{Spectr1}(b) shows the momentum-resolved energy-distribution curves (EDCs) at the $\Gamma$, M, and M$^{*}$ points (using a full window size of 0.2 ${\rm \AA}^{-1}$ for $k_x$ and $k_y$ in the $k$-space) without normalization, with the assumption that the photon intensity in this region remains constant. 
For M/M$^{*}$ points, the data were averaged across three equivalent points in the first Brillouin zone to improve the statistics.
Photoelectron intensity maps at binding energies (BE) of 0.1$\pm$0.1 eV and 2$\pm$0.1 eV are displayed in Figs. \ref{Spectr1}(c) and (d), respectively. 
 
 Figures \ref{Spectr1}(e) shows a photoelectron intensity map where the horizontal axis is $k_x$ along the M-$\Gamma$-M$^*$ line 
 ( $k_y$=0 using a full window size of  0.2 ${\rm \AA}$ ) and the vertial axis shows the BE.
 In Figs. \ref{Spectr1}(c-e), we do not show a detail of the resonant behavior of the photoelectron emission (this will be discussed later) but rather present  
the overall band structures. Then, all the photoelectron intensities recorded at photon energies from I to V have been averaged to enhance the statistics and mitigate the resonance effect in Figs. \ref{Spectr1}(c-e).
The first Brillouin zone and the positions of M and M$^{*}$ points are depicted in Figs. \ref{Spectr1}(c) and (d), respectively.
It is worth mentioning that the notations of M (M$^*$) and $\Gamma$ are not accurate but approximate, as the $k_z$ value is estimated to range from $-0.15$ ${\rm \AA}^{-1}$ ($h\nu$=455 eV) to $-0.10$ ${\rm \AA}^{-1}$ ($h\nu$=459 eV) at the ``M" point and from $-0.1$ ${\rm \AA}^{-1}$ ($h\nu$=455 eV) to $-0.05$ ${\rm \AA}^{-1}$ ($h\nu$=459 eV) at the ``$\Gamma$" point. The determination of the $k_z$ values is based on the assumption as the 
inner potential $V_0$ of 13 eV~\cite{Chen2016}.

In this analysis, we direct our attention towards three prominent peaks: Peak A, originating from the electron pocket crossing the Fermi level at both the M and M$^{*}$ points; Peak B, associated with the hole pocket situated just below the Fermi level at the $\Gamma$ point; and Peak C, corresponding to the band with BE of 2 eV at the M point as shown in Fig. 3(b).
The peak A vanishes at the photon energy below the Ti2p threshold (I) and becomes faintly observable at photon energies II-IV, with particularly heightened intensity at V as clearly recognized in Fig.~\ref{Spectr1}(b).  
The peak A demonstrates roughly similar intensities at both the M and M$^{*}$ points. 
The peak B exhibits a comparable intensity trend to the peak A, as shown in Fig.~\ref{Spectr1} (b).
The peak C, on the other hand, exhibits a somewhat different behavior. It is already observable at the photon energy I (below the Ti2p X-ray absorption threshold), and its intensity does not exhibit any resonant enhancement at the photon energy V where the peaks A and B show clear enhancements in intensity. 
Moreover, the peak C' at the M$^{*}$ point consistently exhibits lower intensity compared to the peak C at the M point across all the photon energies. 
The asymmetry in intensity between the peak C and C' is further highlighted in Fig.~\ref{Spectr1}(d and e).
Considering that M and M$^{*}$ are energetically degenerate in the $k$-space, the asymmetry in photoelectron intensity between them is attributed to the matrix element effect during the photoexcitation process. This discrepancy may arise when one takes into account the light polarization vector, as depicted by the yellow allow in Fig.~\ref{Spectr1}(d).
Figure \ref{Spectr2}(a) shows the CIS spectra of the entire valence band, that mirrors Ti2p$_{3/2}$-to-bound excitation [Fig.~\ref{Spectr1}(a)], together with 
the result of the least-square fitting using two  Lorentzian peaks. The peak positions  (widths) in the photon energy for the Ti3d-$t_{2g}$ and the Ti3d-$e_{g}$ peaks
were estimated 456.9 and 459.3 eV (0.8 and 1.6 eV), respectively.
CIS spectra for the peaks A and B [defined in Fig.~\ref{Spectr1}(b)] are presented in Figs. \ref{Spectr2}(b) and (c), where the photoelectron intensity yields, with window sizes for electron energy and momentum set at 0.2 eV and 0.4 ${\rm \AA}^{-1}$, respectively, are plotted as dots. These spectra cannot  be fitted using two Lorentzian curves as was possible in Fig.~4(a), but two additional components were required to achieve a reasonable agreement with the experimental results [Figs. \ref{Spectr2}(b,c)].
Two components  (blue and red lines) are fixed to the same energies and widths as the CIS spectrum of the total VB (relative intensities serve as fitting parameters), while the other two (dashed lines in green and violet) are allowed to be independently adjusted. The energy positions of additional components are set at 456.3 and 458.7 eV for the peak A at the M point [Fig.~\ref{Spectr2}(b)], and 456.5 and 458.7 eV for the peak B at the $\Gamma$ point [Fig.~\ref{Spectr2}(c)]. 
All the components observed in the CIS spectra of the peaks A and B were represented by the symmetric Lorentzian curves and no asymmetric Fano shaped curve 
was required. As mentioned in the introduction, this suggests that the photoelectron process is largely governed by the cascade process following the core excitation (processes A$\rightarrow$B/B' in Fig. \ref{Fig2})
without an interference with the direct photoelectron process (C in Fig.~\ref{Fig2}). 
 This implies that the probability of direct photoelectron emission processes is considerably low~\cite{Fano1961, Limonov2021}, which is consistent with the very low
 intensities of the peaks A and B in ARPES/MRPES at the photon energy below the Ti2p threshold [Fig.~\ref{Spectr1}(b)].

Two additional peaks, distinct from the two X-ray absorption peaks corresponding to the Ti2p$_{2/3}$ $\rightarrow$ Ti3d-$t_{2g}$ and  Ti2p$_{2/3}$ $\rightarrow$ Ti3d-$e_{g}$ excitations[Fig.~\ref{Spectr2} (a)], are observed in both the CIS of peak A (BE=0.1 eV) at the M-point and peak B (BE=0.2 eV) at the $\Gamma$ point, respectively[Figs. \ref{Spectr2} (b,c)]. These additional peaks in the CIS spectra exhibit shifts of 0.6eV from the Ti3d-$t_{2g}$ peak and 0.5 eV from the Ti3d-$e_{g}$ peak in the XAS spetrum  for the peak A [Fig. \ref{Spectr2} (b]], and 0.5 eV/0.5 eV from the Ti3d-$t_{2g}$/Ti3d-$e_{g}$ peaks ifor the peak B [Fig. \ref{Spectr2} (c]].   
Plausibly, these shifts can be attributed to surface sites (including  the  
surface defect site such as the Se-vacant at the surface), considering the short escape depth of the photoelectrons possessing a kinetic energy of several hundred eV.
This discovery represents, to the best of our knowledge, the first experimental evidence of a surface-energy shift in the X-ray absorption spectrum of the TiSe$_2$ surface.
 It is important to note, for a detailed discussion, that the excitation energy of the core-to-bound state is determined by three factors: a) the energy shift in the Ti2p core-level, b) that in the conduction-band, and c) the shit in the core-exciton binding energy. 

 In recent X-ray photoelectron spectroscopic works on the TiSe$_2$ crystal~\cite{Shkvarin2016,Starnberg_2023},  no surface energy shift 
in the Ti2p core level has been observed on this surface. This  contradicts  the factor (a).  Moreover, there has been no reports on the surface peak in the valence band photoelectron 
spectroscopy~\cite{Stoffe1985,Rossnagel2002,Cercellier2007,Chen2016,Watson2019,Wegner2020},which may suggest that there is no bulk-to-surface energy shift in the conduction band as well. Consequently, the factor (b)  does not seem to be the main reason for the observed phenomenon.
Therefore, it's plausible that the factor c), i.e., the energy-shift due to the formation of the  surface core exciton, dominates the surface energy-shift in the CIS spetra of TiSe$_2$.
Although the surface core exciton  has seldom been investigated~\cite{Ichikawa1991,TANAKA1996,TANAKA_Xenon2001},  it is crucial to  be accounted, especially when a significant Coulomb interaction may be expected between the core-hole and the bound electron in the conduction band. A recent study on the CaF$_2$ surface proposed the surface shift in the exciton binding energy using sophisticated DFT calculations employing a many-body approach~\cite{Matias2022}.
Our finding implies a possible difference in the excitonic interaction between the Ti2p core hole and the Ti3d conduction band at the surface and the bulk.
We note that the intensity ratios between these additional peaks and those in the X-ray absorption spectrum differ substantially in Ti3d-$t_{2g}$ and Ti3d-$e_{g}$ peaks (peaks around 456.9 and 459.3 eV, respectively) both in spectra in Fig.~\ref{Spectr2} (b) and (c).
This divergence indicates that a simple assumption of a single site with a single energy cannot  explain the spectrum. A more intricate scenario, involving multiple reconstructions and/or defects at the surface, should be considered.
A more systematic study, including complementary data regarding the surface condition of the sample, is required to reveal 
the true nature of the excitonic interaction at the TiSe$_2$ surface. This aspect will be investigated separately in the near future after improving the energy resolution
of the instrument.

Meanwhile, Figs.~\ref{Spectr2} (d) and (e) show the photoelectron intensity maps as functions of the photon energy and the binding energy at the M and M$^*$ points, respectively. Figures \ref{Spectr2} (d) and (e) illustrate the photoelectron intensity maps as functions of photon energy and binding energy at the M and M$^*$ points, respectively. Notably, only at the M point[Fig.~\ref{Spectr2} (d)], the intensity at BE=2 eV is evident at all photon energies, consistent with the EDC spectra [Figs.\ref{Spectr1} (b,d,e)].
The CIS spectra of the peaks C and C' in Fig.~\ref{Spectr1} (b), along with their difference, are displayed in Fig.~\ref{Spectr2}(f), utilizing the same window sizes as those for the peaks A and B in Figs. \ref{Spectr2}(b and c). 
In both CIS spectra of the peaks C and C'[Fig.~\ref{Spectr2}(f)], two peak structures are observed around  456.9 eV and 459.3 eV (indicated by the dotted lines) similarly to the the X-ray absorption spectrum [Fig.~\ref{Spectr2}(a)]. However, these peak structures may not correspond to  the true intensity changes in the peak C but rather to changes in the background. 
In the EDC spectra presented in Fig.~\ref{Spectr1}(b),  the peaks C and C' are situated on the shoulder of the peak centered around BE=3 eV, which exhibit a strong enhancement at
the photon energy of 459.3 eV as shown in the EDC spectrum V in Fig.~\ref{Spectr1} (b), and this enhancement
may be contributing much to the apparent resonant enhancement of the  peaks at 456.9 eV and 459.3 eV in the CIS spectra of 
the peaks C and C' [Fig.~\ref{Spectr2} (f)].
The intensity of the peak at BE$\sim$3eV, probably contributed mainly from the Ti3d state, remains relatively consistent from M to M$^{*}$, as also depicted in Figs. \ref{Spectr2}(d) and (e). 
Consequently, 
their difference shown in Fig.~\ref{Spectr2}(f) can be interpreted as an approximate true yield spectrum, effectively isolated from the background due to the Ti3d-related electron emission, of the peak C.
It should be noted  that no adjustments in intensities between the CIS spectra for the peaks C (at M) and C' (at M$^{*}$) were carried out before the subtraction
except averaging the intensity at three equivalent M and M$^*$ points.
The photon-energy dependence of the peak C is drastically different from the situation of the peaks A and B, where the resonance enhancements in symmetric Lorentzian 
curves were observed at the Ti2p-to-bound excitation photon energy as shown in Figs. \ref{Spectr2} (b and c).
In contrast, it is clear that almost no ``resonant enhancement" in the photoelectron emission is observed for the peak C  in  the excitation photon energy region between 455 and 458 eV as seen in Fig.~\ref{Spectr1}(b). 
Instead, the largest peak intensity of the peak C is achieved when the photon energy is just below  the Ti2p photoexcitation threshold [case I in Fig.~\ref{Spectr1} (b)],
and become smaller when the photon energy passed over  456.9 eV and 459.3 eV, which correspond to the excitation from Ti2p$_{3/2}$ to Ti3d-$t_{2g}$ and Ti3d-$e_{g}$  [Fig.~\ref{Spectr2}(f)]. 
 The disparities between the CIS spectra at M and M$^{*}$ points can be fitted using non-conventional Fano-type asymmetric peak shapes. In Fig. \ref{Spectr2}(g), 
 two Fano-shaped curves (eqs.~\ref{Fano_formula} and  \ref{Fano_energy}) are employed for fitting. In this fitting, $x_0$ (456.9 and 459.3 eV) and $\Gamma$ (0.8 and 1.6 eV) are kept consistent with the values in the CIS spectrum of the entire VB [Fig.~\ref{Spectr2}(a)], and  only two sets of parameters  of $I_0$ and $q$ were adjusted. The results exhibit a remarkable alignment with the experimental outcomes.
 A noteworthy emphasis lies in the fact that the obtained $q$ parameters are both negative, with values of $-0.8$ and $-0.5$, a phenomenon hitherto unobserved in resonant photoelectron spectroscopy to the best of our knowledge.
  
 \begin{figure}[bthp]
  \centering
\includegraphics[width=9cm]{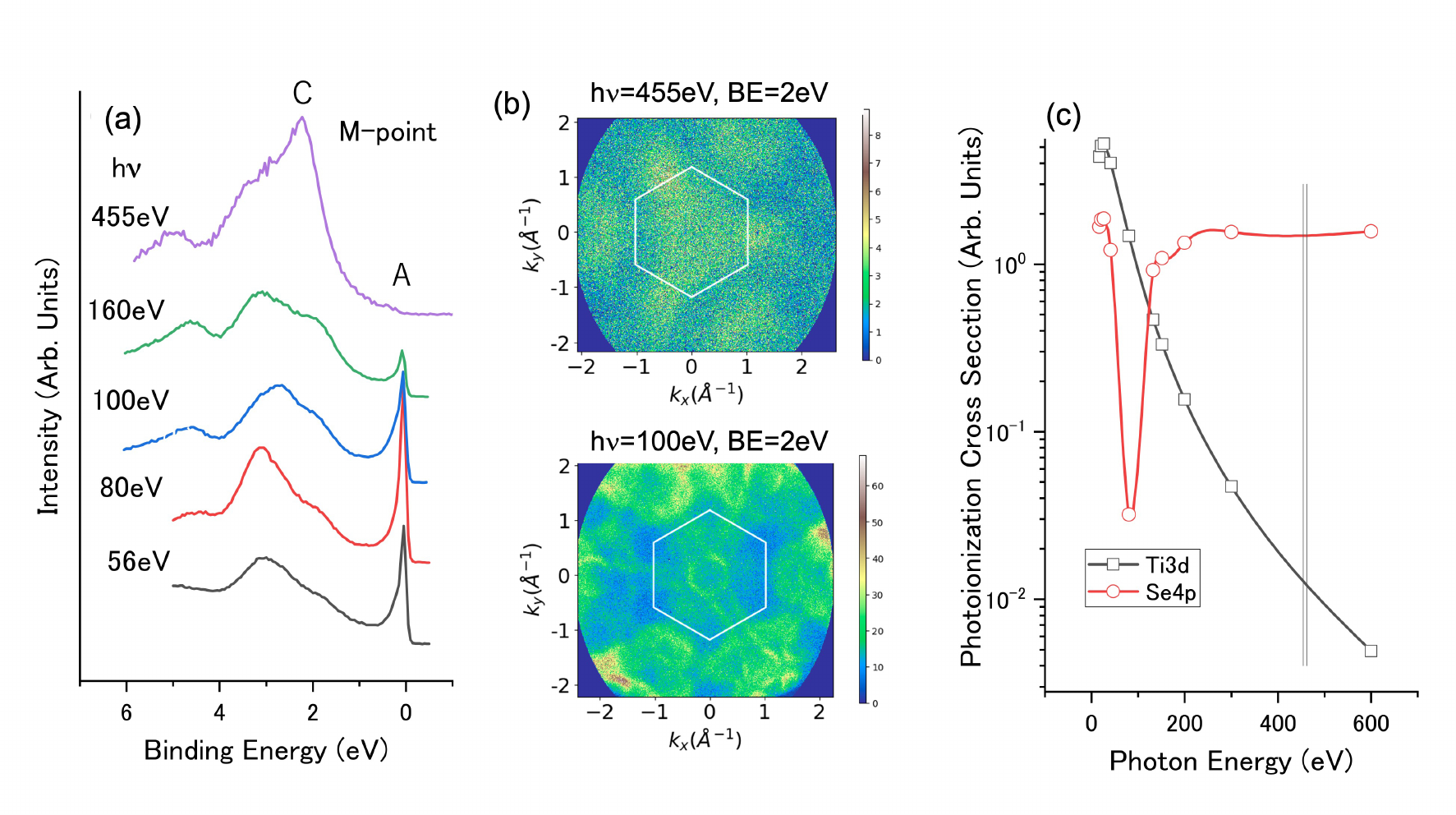}
  \caption{ \label {hvdep} (a) Momentum-resolved photoelectron spectrum at M(L) point taken at the photon energy indicated.
   (b) Two-dimensional photoelectron intensity maps at BE=2 eV taken at h$\nu=$455 eV (upper map) and 100 eV (lower map), 
   (c) Calculated photoionization cross section for Ti3d and Se4p orbitals~\cite{YEH19851}. 
  }
\end{figure}

For  more comprehensive exploration of the photoelectron emitting process and the peak interpretation, a study encompassing a wider range of photon energies was conducted. Figure \ref{hvdep}(a) presents EDC profiles at the M point (excluding M$^{*}$) across several photon energies. Each spectrum is normalized in intensity to the  averaged amplitude in the observed binding energy range. 
Meanwhile, Fig. \ref{hvdep}(b) displays an iso-energy photoelectron intensity maps at BE= 2 eV,
recorded at photon energies of 455 and 100 eV.
 
 A clear observation emerges: the peak A near the Fermi level exhibits pronounced strength at lower photon  energies below 200 eV but disappears drastically  at $h\nu$=455 eV (note that this value lies below the Ti2p threshold which starts at around 456 eV). Conversely, the peak C at BE=2 eV is almost absent at lower photon energies below 100 eV but becomes prominent at 
 $h\nu$=455 eV, both in the M-point EDC and the iso-energy intensity maps at BE=2 eV [Figs. \ref{hvdep} (a,b)]. This divergence can be attributed to the varying photoionization probability of atomic orbitals as a function of the photon energy.
Figure \ref{hvdep} (c) displays the photoionization cross section of the Ti3d and Se4p orbitals, as calculated by Yeh and Lindau~\cite{YEH19851}. Notably, the Se4p cross section exhibits a prominent dip around $h\nu$=100 eV and remains relatively constant from 200 eV to 600 eV. Conversely, the Ti3d cross section decreased gradually and monotonically with the photon energy up to 600 eV. The narrow thin lines indicate the photon energy range (455-459 eV) of our resonant photoelectron spectroscopy measurements. Within this range, the photoionization cross section of the Ti3d state  decreases significantly down to  approximately 1\%  of that of the Se4p state. Consequently, (direct) photoelectron emission from the bands predominated by the Se4p orbital occurs at the photon energies within 455-459 eV, while the dominance of the Ti3d orbital is evident at photon energies at $\sim$80-100 eV. As a result, the peaks A and B in Fig.~\ref{Spectr1}(b) can be attributed to the Ti3d-derived band, whereas the peak C corresponds to the Se4p-derived band.

 \begin{figure}[bthp]
  \centering
\includegraphics[width=8cm]{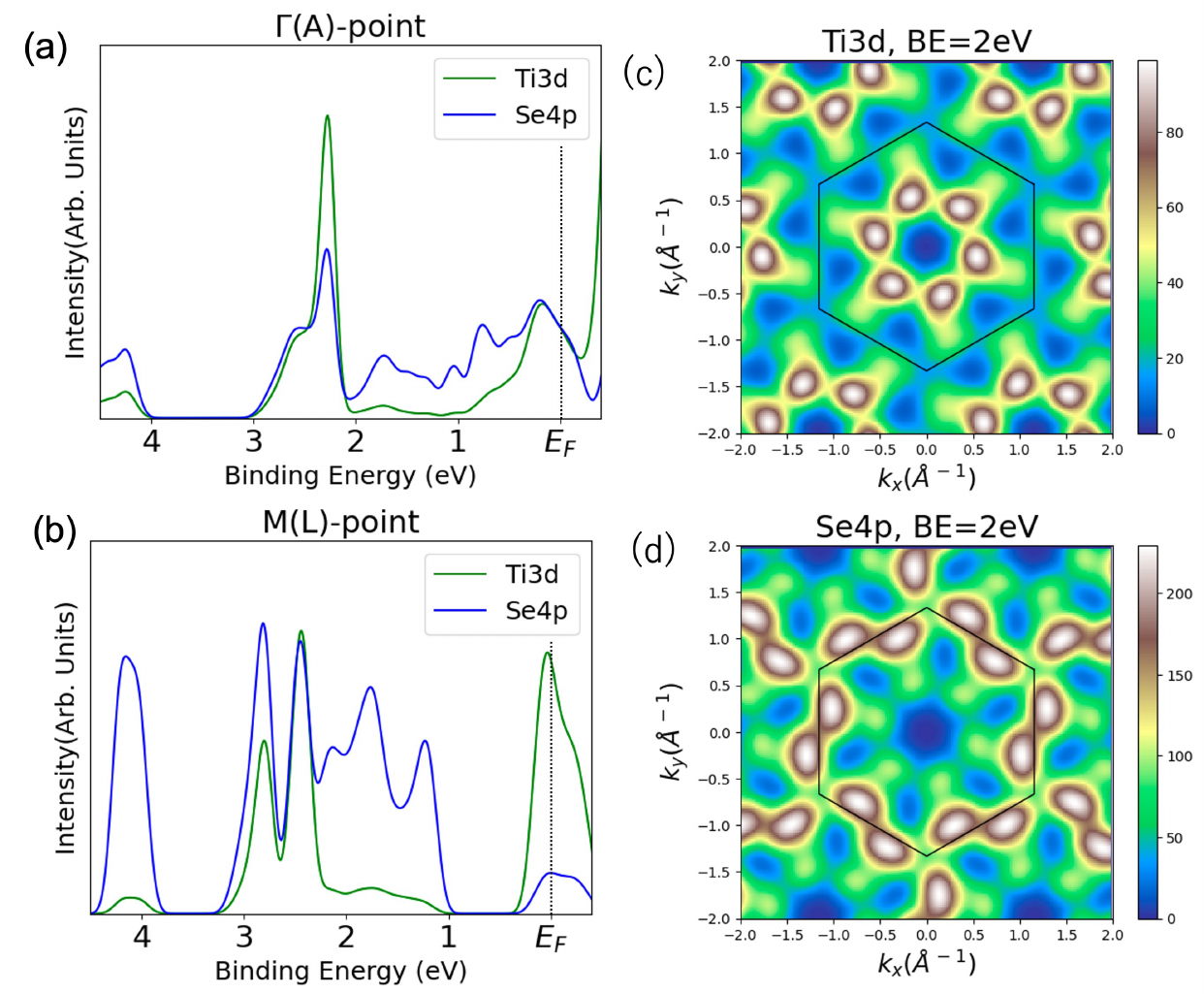}
  \caption{  \label {DFTCalc} Results of the DFT calculations: (a,b) Projected energy-distribution curves for the Ti3d  and Se4p  orbitals at $\Gamma$-A (a) and  
  M-L (b) points. The density of states are integrated along the $k_z$ line in the Brillouin zone. (c,d) Projected density of states
  of Ti3d(c) and Se4p(d) orbitals in the $k_x$-$k_y$ space at BE=2 eV. }
 \end{figure} 

To validate the aforementioned interpretation, we conducted Density Functional Theory (DFT) calculations, which yielded wavefunction projections onto atomic orbitals. The populations of Se4p and Ti3d orbitals in bands were determined using a $64 \times 64 \times 12$ $k$-mesh. Convolutions were applied in both momentum and energy, utilizing integration window sizes of $\Delta k = 0.1$ ${\rm \AA}^{-1}$ and $\Delta E = 0.1$ eV, respectively, across the Brillouin zone. The subsequent results include EDC at specific $k$-points and iso-energy density of states at particular binding energies. 

Figures \ref{DFTCalc}(a) and (b) respectively illustrate the intensity distribution of  the Ti3d and Se4p bands at the $\Gamma$(A) and  M(L) points.  
 The calculated outcomes are integrated along the $k_z$ axis within the Brillouin zone, since there is  a non-negligible dispersion along the $k_z$ axis in some bands (refer to Fig. 1(a)) as the exact energy position of the band is not expounded and is not important here.
On analyzing the $\Gamma$ (A) point [Fig.~\ref{DFTCalc}(a)], it becomes evident that the states around the Fermi level have the rather equivalent contributions from the Se4p and Ti3d orbitals, although the absolute binding energy is somewhat shifted to higher binding energy from the experimental result[Figs. 1(a,b) and Fig. 3(b)]. 
This  calculation signifies the  hybridization between the Ti3d and Se4p orbitals at the top of the valence band at the $\Gamma$ (A) point. 
This correlates well with the CIS spectrum for the peak B [Fig.~\ref{Spectr2} (c)], where the resonant enhancement in the photoelectron intensity is evident at the photon energy corresponding to the Ti2p$\rightarrow$Ti3d excitation indicating the valence band top (the peak B) is contributed from the Ti3d orbital. 
It is noteworthy that the hybridization of the valence band between the Ti3d and Se4p states at the $\Gamma$(A) point in the CDW phase has been previously postulated~\cite{Stoffe1985, Rossnagel2002, Hellgren2017, Watson2019}, while our experiments and the DFT calculation save been made on  the non-CDW phase. 
The Ti3d orbital contributes dominantly to the states crossing the Fermi level at the M (L) point, as shown in Fig.~\ref{DFTCalc}(b). 
This is consistent with the CIS spectrum of the peak A at the M (L) point in Fig.~\ref{Spectr2} (b).

Conversely, the electronic states of BE=1-2 eV at the M (L) point [Fig.~\ref{DFTCalc}(b)] are predominated by the Se4p orbital, aligning with the interpretation as stated above. This is further substantiated by comparing the calculated results of the projected iso-energy two-dimensional density of states at the binding energy of 2 eV (Figs. \ref{DFTCalc}(c) and (d)) to the photoelectron intensity map taken at h$\nu$=455 eV (Se4p-dominated) and 100 eV (Ti3d-dominated)[Fig.~\ref{hvdep}(b)]. These photon energies are sensitive to Ti3d and Se4p, respectively, as predicted in Fig. 4(c).
Though the detailed structures might not be in perfect agreement, possibly due to dispersion along the $k_z$ axis, the overarching trend persists. Specifically, the Ti3d states predominate in the vicinity of the $\Gamma$ (A) point, while the Se4p states prevail around the M (L) point. 

\begin{figure}[bthp]
  \centering
\includegraphics[width=8cm]{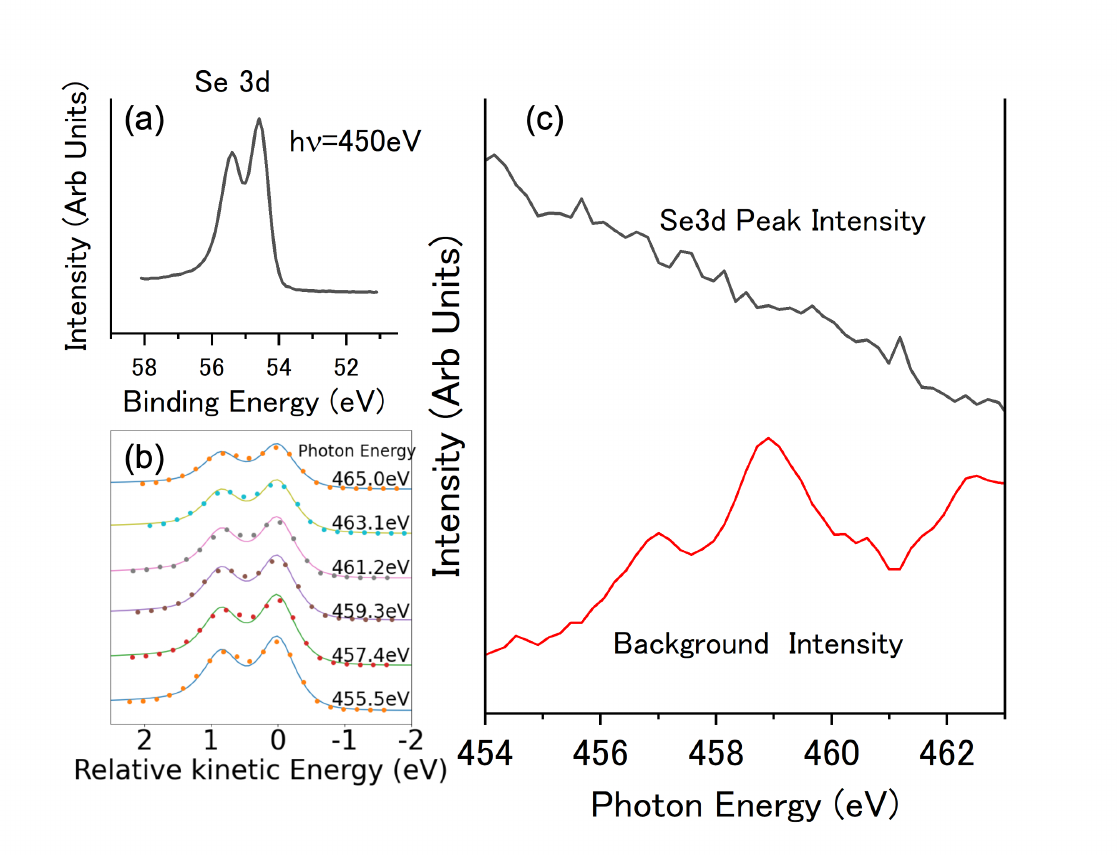}
  \caption{  \label {MARPECheck}  (a) Angle-integrated photoelectron spectrum recorded at $h\nu$=450 eV in the region of the Se3d core level.
  (b) Series of the photoelectron spectra with changing the photon energy. The dots show the actual data and the lines show the results of the peak fitting using
  the Voigt peak shape and linear background. (c) Plots of the peak intensity and background intensity as a function of the photon energy. }
 \end{figure}

In accordance with the interpretation of the peak C being predominated by the Se4p orbital, its CIS spectrum [Fig.~\ref{Spectr2}(g)] can be attributed to an Auger electron emission originating from the same band subsequent to the Ti2p$\rightarrow$Ti3d excitation.  To further discuss the detail of this Auger electron emission proccess, two distinct possibilities have to be checked.
The first possibility is the CVV Auger decay, wherein C denotes the core level (Ti2p), and the two V's represent the valence bands (Ti3d and Se4p). This scenario is depicted in Fig.~\ref{Fig2}. Alternatively, a cascade (intra-atomic) CVV Auger decay may transpire, where all pertinent levels (bands) are derived from the Se states following an inter-atomic CCC (Ti/Se/Se) Auger decay  which yields the core hole at the Se atom.
This inter-atomic CCC Auger process, in which  the Se-core holes are provided after the de-excitation of the Ti2p core hole, is the same as the concept of the previous MARPE.~\cite{Kay1998, PhysRevB.63.115119, FUJIKAWA2022147202}
In the latter process, the Fano line shape in the CIS spectrum of the peak C [Fig.~\ref{Spectr2}(g)] is predominantly determined by the conventional core-level MARPE process. To investigate this possibility, angle-integrated photoelectron monitoring of the Se3d core level was conducted while varying the photon energy around the Ti2p$\rightarrow$Ti3d excitation threshold. Figure \ref{MARPECheck}(a) illustrates the Se3d peak recorded at $h\nu$=450 eV, and Fig.~\ref{MARPECheck}(b) showcases examples of the Se3d peak spectra at varying photon energies. The dots in the graphs depict actual data, while the lines represent the results of a least-square peak fitting calculation employing the Voigt function together with the Shirley background function alongside a linear background. The intensities of the peak area and the averaged background are plotted against the photon energy in Fig.~\ref{MARPECheck}(c).
The outcome is clear: Despite the enhancement in background intensity, which likely arises from secondary electrons, there is no resonance in Se3d photoelectron emission near the Ti2p$\rightarrow$Ti3d excitation. In essence, we do not observe core-level MARPE in TiSe$_2$, and can deny the cascade process. Then, it leads to the conclusion that the Fano resonance observed in Fig.~\ref{Spectr2}(g) arises from the interference between the Se4p photoelectron emission and the inter-atomic $CVV$ (Ti2p/Ti3d/Se4p) Auger electron emission. 

\section{Discussion}

It appears evident that both the conduction band bottom at the M(L) point and the valence band top at the $\Gamma$(A) point [the A and B peaks
in the EDC spectra shown in Fig.~\ref{Spectr1}(b)]
are contributed by the Ti3d orbital. This conclusion is supported by the noticeable increases in photoelectron intensities 
at the photon energy corresponding to the Ti2p$\rightarrow$Ti3d photoexcitation [shown in Figs.~\ref{Spectr2}(b,c)]. 
This interpretation gains further support from: a) A comparison of the photon energy dependence between ARPES/MRPES 
and the cross-section for photoionization of atomic orbitals [depicted in Figs. \ref{hvdep}], and b) Orbital-specific DFT calculations 
 [illustrated in Figs. \ref{DFTCalc} (a, c)].
 The CIS spectra shown in Figs. \ref{Spectr2}(b, c) do not consist of the asymmetric Fano-type curve that has been commonly
 observed in resonant photoelectron spectroscopy~\cite{Davis1986,Kakizaki1983,Kotani2001}, but consist of
 symmetric Lorentzian curves. This discrepancy can be explained by considering the very weak Ti3d photoionization cross-section at the photon energies of around 455 eV [Figs. 4(c) and 5(c)].
 As mentioned already in the introduction, the conventional resonant photoelectron emission process is induced by the interference between the direct photoelectron emission process [process C in Fig.~\ref{Fig2} ] and the  Auger electron emission following the core-hole creation [processes A and B/B' in Fig.~\ref{Fig2} ].
Within the photon energy range corresponding to the Ti2p$\rightarrow$Ti3d photoexcitation , the photoionization cross-section of the Ti3d states is extremely small compared to that of the Se4p states, leading to the predominance of the Auger electron emission. 
Consequently, effective interference does not occur, resulting in infinite value of the parameter $q$ in the  Fano resonance.
As a consequence, the symmetric Lorentzian curve is achieved as shown in Figs.~\ref{Spectr2} (b,c).

The evaluation of the contribution from the Se4p orbital through the Ti2p-resonant photoelectron spectroscopy is not easy  
when the observed band has a significant contribution from the Ti3d orbitals which can overshadow the weak resonance of the Se4p state. 
Therefore, our present results do not discount the possibility of hybridization between the Ti3d and Se4p orbitals.
Concerning the top of the valence band [peak B in Fig.3(b)], one expects the contribution from the Se4p state considering the 
simple electronic configuration of TiSe$_2$ as (Se4p)$^6$(Ti3d)$^0$. 
Consequently, at the top of the valence band at the $\Gamma$(A) point, the Ti3d-Se4p hybridization is highly likely. 
It's worth noting that the Ti3d-Se4p hybridization at the top of the valence band has been previously proposed as 
a result of the CDW transition, wherein the M(L) point is folded into the $\Gamma$(A) point~\cite{Stoffe1985,Rossnagel2002,Hellgren2017,Watson2019}.
 However, our current findings suggest that this hybridization occurs even in the absence of the CDW phase. 
Theoretical work based on our experimental results is highly expected to elucidate the detailed mechanism of the CDW transition.
 
Meanwhile, the band corresponding to the peak C (BE$\sim$2 eV at the M-point) in ARPES/MRPES spectra is predominated 
by the Se4p orbital, as proved by the photon energy dependent ARPES/MRPES and DFT calculations. 
In conventional resonant photoelectron spectroscopy, 
the resonant enhancement occurs for the photoelectron emission from  a band consisting of the core-excited atom's orbitals,
and a band originating from an orbital not associated with the core-excited atom is typically assumed to have no significant resonance.
However, in the case of the peak C [BE$\sim$2 eV at the M point in Fig. 3(b)], the CIS spectrum, after subtracting the CIS spectrum of C' (at the M$^*$ point) for the background elimination, can be successfully fitted using the Fano-resonance function as shown in Fig. 4(g) . This indicates that there is some interference between the photoelectron emission process from the Se4p-derived band and the inter-atomic Auger electron emission following the Ti2p core-level excitation[Figs. 2].
On the ``exchange" Auger decay (process B in Fig.2), the transition from the excited Ti3d state to the Ti2p core hole  provides the energy for the electron emission from the occupied Se4p state via the inter-atomic Coulomb interaction. 
In the case of the ``direct" Auger decay (process B' in Fig.2), meanwhile, the inter-atomic transition from the Se4p state to the Ti2p core hole provides 
the energy for the electron emission from the Ti3d state. These ``direct'' and ``exchange'' Auger processes are energetically degenerate.
It is important to note that the inter-atomic Coulomb interaction is generally much weaker than the intra-atomic interaction. 
As a result, the inter-atomic resonance effect is expected to be much smaller compared to the intra-atomic resonance effect observed in the conventional resonant photoelectron spectrum.
This likely explains why such an effect has not been previously reported. In the current scenario on TiSe$_2$, however, the direct photoelectron emission process from the Ti3d 
state is much weaker than that for the Se4p sate[Fig. 4(c)]. 
This  is the reason why the  Se4p resonance can be observed without being overshadowed by the Ti3d resonance in the present case.

Another important result observed for the first time in this research is  the negative $q$ value in the Fano resonance in resonant photoelectron spectroscopy. 
Although negative $q$ has often been reported  in many researches (in fact, negative $q$ was reported in the very first application of 
the Fano formula to the  $2s2p^1P$  autoionized level at $\sim$60 eV of double excitation of the He atom~\cite{Fano1961}),  only
positive $q$ has been reported in the resonant photoelectron spectra~\cite{Davis1986,Kakizaki1983,Kotani2001}. 
In analogy to classical mechanics~\cite{Joe_2006,Iizawa_2021}, Fano resonance can be described as
an interference between two coupled harmonic oscillators with different frequencies and damping factors that determine the spectral width.
The damped oscillator with a wider spectral width corresponds to the  continuous levels, and the oscillator with a narrow width
corresponds to the distinct levels in the ordinal Fano resonance.
The parameter $q$ is determined by  the phase shift of the damped oscillator at the resonant frequency. 
The sign of $q$ depends on the sign of the frequency difference between the two oscillators~\cite{Joe_2006,Iizawa_2021}. 
In the quantum mechanics, meanwhile, $q$ is also connected to $\phi$ which is a phase-shift between two excitations (distinct and continuous) by $q=-\cot \phi$~\cite{Ott2013,Limonov2021}.
Furthermore, it has been reported that  $q$ can be controlled from negative to positive by tuning the temporal condition between
the distinct and continuous transitions using the ultrashort pulse lasers~\cite{Ott2013}. 
Therefore, it is considered that the sign of $q$ in the resonant photoelectron emission 
 is determined by the temporal characteristics  of the core-excitation/Auger decay and the direct photoelectron emission processes.
A fundamental distinction of the current situation, marked by a negative $q$, from the conventional resonant photoelectron emission lies in the contrast between inter-atomic and intra-atomic interactions during  the Auger decay. It seems reasonable to posit that these processes exhibit diverse phase conditions concerning the interference between direct photoelectron emission and Auger electron emission. 
It seems interesting to point out that  only  positive $q$ values have been reported in the resonant photoelectron emission  associated with the inter-atomic core-core-core Auger decay known as MARPE~\cite{Kay1998, PhysRevB.63.115119, FUJIKAWA2022147202}.
The discernible contrast between conventional MARPE and our current investigation suggests a variance in the dynamics of excitation-decay during photoelectron/Auger electron emission. This discrepancy becomes apparent when comparing scenarios involving solely localized core states to those where delocalized valence states are involved.
To gain deeper insights into the sign of $q$ during resonant photoelectron emission, a more intricate theoretical investigation focusing on the microscopic dynamics of excitation and decay is required.
Furthermore, conducting similar experiments on magnetic compounds excited by circularly polarized synchrotron radiation light could potentially unlock avenues for a more comprehensive discussion of inter-atomic interactions, encompassing spin-spin and spin-orbit couplings, within resonant photoelectron spectroscopy.

\section{Conclusion}
In this study, we have conducted an investigation by means of momentum-resolved resonant photoelectron spectroscopy on 1T-TiSe$_2$ at the Ti2p-edge, focusing under room temperature conditions. 
This investigation was made possible by the use of high performance momentum microscopy equipment. 
The investigation was further facilitated by analyzing the comprehensive range of photon-energy dependence within the momentum-resolved photoelectron spectra. This analysis was complemented by the use of theoretical results of the atomic orbital photoionization cross sections, together with density functional theory (DFT) calculations.
Through the combined efforts of these experimental and theoretical approaches, we have uncovered the following significant findings:
\begin{enumerate}
\item The valence band located at the $\Gamma$(A) point exhibits the intricate hybridization involving both the Ti3d and Se4p orbitals, even at room temperature. Notably, this hybridization is not a consequence of the charge-density-wave (CDW) transition.

\item The CIS spectrum for the Ti3d-derived band at the Ti2p edge is not characterized by the Fano-type curve, but by the symmetric Lorentzian curve. This can be understood by the fact that the photoionization cross section of the Ti3d orbital is quite small at the photon energies corresponding to the Ti2p edge. Therefore, the photoelectron emission in this photon energy region is almost dominated by the Auger electron emission following the excitation from Ti2p to Ti3d states, and the Fano interference is effectively suppressed.

\item The band primarily characterized by the Se4p orbital, located at the binding energy of $\sim$2 eV at the M-point, exhibits a fascinating resonant phenomenon. This peculiar behavior is evident through the manifestation of a distinct Fano line shape featuring a negative value of $q$. The occurrence of Fano resonance within the Se4p-derived band can be attributed to the intricate interference between the direct photoelectron emission originating from this band and the inter-atomic $CVV$ (Core-Valence-Valence, {\it i.e.}, Ti2p/Ti3d/Se4p) Auger electron emission process. 
Notably, this observation of a negative $q$ value is the first  occurrence in the field of resonant photoelectron spectroscopy. 
It is also noteworthy that the experimental evidence has demonstrated the unique nature of this result of the electron emission from the valence band in contrast to the conventional multi atom resonant photoemission (MARPE), where only core-level states are involved in inter-atomic Auger electron emission.\end{enumerate}

\section{Acknowledgement}
We would like to thank UVSOR staff for their kind support for the experiments.
This work was performed at the BL6U of UVSOR Synchrotron Facility, 
Institute for Molecular Science (IMS program: 20-212, 21-201). 
The computation was performed using Research Center for Computational Science, Okazaki, Japan (Project: 21-IMS-C161,22-IMS-C162).
This work was supported by JSPS KAKENHI Grant Numbers 21K04826 and 22H05445.

\%apsrev4-2.bst 2019-01-14 (MD) hand-edited version of apsrev4-1.bst
\end{document}